\documentstyle[12pt]{article}

\title{\bf Quantum lattice fluctuations in a model electron-phonon system}
\author{Q.Wang{\footnotemark[1]} {} {} and {} H.Zheng\\ 
     \footnotesize\em
     Department of Applied Physics, A9707092, Shanghai Jiao Tong University, 
      Shanghai 200240, P.R.China\\ 
     \footnotesize and \em 
     Physics Department, Xinjiang Normal University, Uromuq 830053, P.R.China}
\footnotetext[1]{E-mail: hzheng@online.sh.cn}
\date{}
\begin{document}
\maketitle
\begin{abstract}
An analytical approach, based on the unitary transformation method, has been 
developed to study the effect of quantum lattice fluctuations on the ground  
state of a model electron-phonon system.
To study nonadiabatic case, the Green's function method is used to implement 
the perturbation treatment. The phase diagram and the density of states of        
fermions are obtained. We show that when electron-phonon coupling constant 
$\alpha^{2}/K$ decreases or phonon frequency $\omega_{\pi}$ increases the 
lattice dimerization and the gap in the fermion spectrum decrease gradually. 
At some critical value the system becomes gapless and the lattice dimerization 
disappears. The inverse-square-root singularity of the 
density of states at the gap edge in the adiabatic case disappears because of 
the nonadiabatic effect, which is consistent with the measurement of optical
conductivity in quasi-one-dimensional systems.\\
\newline
{\em PACS\,}: 71.45.Lr, 71.20.-b, 64.70.-p, 63.20.Kr\\
{\em Keywords\,}: quantum lattice fluctuation; nonadiabatic effect;
                          electron-phonon interaction; density of states
\end{abstract}
\newpage
\section{Introduction}
  
  The physical and chemical properties of materials with a
quasi-one-dimensional charge-density-wave (CDW) state, for example, the
halogen-bridged mixed-valence transition-metal complexes and the
conducting polymers, have been the subject of intense study in recent
years, because of their intrinsically interesting properties
and important technological applications \cite{refa1,refa3}.  
Among the models for one-dimensional systems the Holstein
model\cite{refb2} and Su-Schrieffer-Heeger (SSH) model\cite{refa19}
are the two typical electron-phonon
coupling Hamiltonian studied by many previous authors. The Holstein
model is for the on-site electron-phonon interaction with 
dispersionless phonons coupled with electron density operator, while
the SSH model is for the on-bond electron-phonon interaction.

In this work we deal with another one-dimensional
electron-phonon coupling model, the simplified one band Hamiltonian
for the halogen-bridged mixed-valence transition-metal complexes
\cite{refa7,refa8,refa9},
\begin{eqnarray}
 H & = & \sum_{l}\left(\frac{1}{2M}P^{2}_{l}+\frac{1}{2}Ku^{2}_{l}\right)
         -\sum_{l,s}t_{0}(c^{\dag}_{l,s}c_{l+1,s}+c^{\dag}_{l+1,s}c_{l,s})
          +\sum_{l,s}\alpha(u_{l}-u_{l+1})c^{\dag}_{l,s}c_{l,s},
\label{eq:hamilt}
\end{eqnarray}
where $c^{\dag}_{l,s}$ and $c_{l,s}$ are the creation and annihilation 
operators of electrons at site $l$ with spin $s$, 
$u_{l}$(with conjugated momentum $p_{l}$) is the displacement of the $l$
ion, $t_{0}$ is the supertransfer energy of electrons between 
neighboring two orbitals, $\alpha$ is the electron-phonon 
coupling constant, $K$ is the elastic constant and $M$
the mass of ions (throughout this paper we set $\hbar=k_{B}=1$).
We will show that, at least for the spinless case, this model is
the same as the Holstein model in the adiabatic limit
($M\rightarrow\infty$) but may be similar to the SSH model in the
antiadiabatic limit ($M\rightarrow 0$).

Within the adiabatic mean-field approximation, that is,
treating the phonon degrees of freedom classically, the model
can be solved easily. In the half-filled-band case, the system
undergoes a Peierls instability and the ground state is dimerized
with an energy gap at the Fermi points $k=\pm\pi/2$\cite{refb3}.
However, this approach is questionable and it has been shown that
the quantum lattice fluctuations must be taken into account
to satisfactorily describe some physical properties of
quasi-one-dimensional systems\cite{refb4}.
Generally speaking, the quantum lattice fluctuations may 
decrease the CDW order parameter\cite{refa4,refa5}. As the density
of states (DOS) is concerned, the results of adiabatic
approximation have inverse-square-root singularity at the gap edge. 
By considering the quantum lattice fluctuations, the singularity 
at the gap edge may disappear\cite{refa6}. The relationship between Peierls 
distortion and phonon frequency in the range from $\omega_{\pi}=0$ to 
$\omega_{\pi}\rightarrow\infty$ should be studied for understanding the 
physical properties of electron-phonon interactions in nonadiabatic case. 

In Hamiltonian (\ref{eq:hamilt}) the operators of lattice modes, $u_{l}$  
and $p_{l}$, can be expanded by the phonon creation and annihilation
operators $b^{\dag}_{-q}$ and $b_{q}$, and
after Fourier transformation the $H$ becomes
\begin{eqnarray}
 H & = & \sum_{q}\omega_{\pi}\left(b^{\dag}_{q}b_{q}+\frac{1}{2}\right) 
         + \sum_{k,s}\epsilon_{k}c^{\dag}_{k,s}c_{k,s}
          +\sum_{q,k,s}\frac{1}{\sqrt{N}}g(q)(b_{q}+b^{\dag}_{-q})
         c^{\dag}_{k+q,s}c_{k,s},
\end{eqnarray}
where $\epsilon_{k}=-2t_{0}\cos k $ is the bare band function,
$N$ is the total number of sites.    
The dispersionless phonon frequency $\omega_{\pi}=\sqrt{K/M}$ and the
coupling function $g(q)=\alpha\sqrt{1/(2M\omega_{\pi})}[1-\exp(iq)]$.
  
\section{Effective Hamiltonian}  

  In order to take into account the fermion-phonon correlation, an unitary
transformation is applied to $H$,
\begin{equation}
 H'=\exp(S)H\exp(-S),
\end{equation}
where the generator $S$ is
\begin{eqnarray}
 S & = & \frac{1}{\sqrt{N}}\sum_{q,k,s}\frac{g(q)}{\omega_{\pi}}
          (b^{\dag}_{-q}-b_{q})\delta(k+q,k)c^{\dag}_{k+q,s}c_{k,s}.
\end{eqnarray}
Here we introduce a function $\delta(k^{'},k)$ which is a function of the 
energies of the incoming and outgoing fermions in the fermion-phonon 
scattering process. We divide the
original Hamiltonian into $H=H^{0}+H^{1}$, where $H^{0}$ contains the first 
two terms of $H$ and $H^{1}$ the last term. Then the unitary transformation
can proceed order by order,
\begin{equation}
 H'=H^{0}+H^{1}+[S,H^{0}]+[S,H^{1}]+\frac{1}{2}[S,[S,H^{0}]]+O(\alpha^{3}).
\end{equation}
The first-order terms in $H^{'}$ are
\begin{eqnarray}
 H^{1}+[S,H^{0}] & = & \frac{1}{\sqrt{N}}\sum_{q,k,s}g(q)(b^{\dag}_{-q}+b_{q})
                        c^{\dag}_{k+q,s}c_{k,s}\nonumber\\
                 & - & \frac{1}{\sqrt{N}}\sum_{q,k,s}g(q)\delta(k+q,k)
                       (b^{\dag}_{-q}+b_{q})c^{\dag}_{k+q,s}c_{k,s}\nonumber\\
                 & + & \frac{1}{\sqrt{N}}\sum_{q,k,s}
                        \frac{g(q)}{\omega_{\pi}}\delta(k+q,k)
                        (b^{\dag}_{-q}-b_{q})(\epsilon_{k}-\epsilon_{k+q})
                        c^{\dag}_{k+q,s}c_{k,s}.
\end{eqnarray}
Note that the  
ground state $|g_{0}\left.\right\rangle$ of $H^{0}$, the non-interacting 
system, is a direct product of a filled fermi-sea $|FS\left.\right\rangle$ 
and a phonon vacuum state $|ph,0\left.\right\rangle$\cite{refa14}:
\begin{equation}
|g_{0}\left.\right\rangle=|FS\left.\right\rangle|ph,0\left.\right\rangle.
\end{equation}
Applying the first-order terms on $|g_{0}\left.\right\rangle$ we get 
\begin{eqnarray}
 (H^{1}+[S,H^{0}])|g_{0}\left.\right\rangle 
       & = & \frac{1}{\sqrt{N}}\sum_{q,k,s}g(q)b^{\dag}_{-q}
             c^{\dag}_{k+q,s}c_{k,s}\left[1-\delta(k+q,k)\left(1-
             \frac{\epsilon_{k}-\epsilon_{k+q}}{\omega_{\pi}}\right)\right]
             |g_{0}\left.\right\rangle,
\end{eqnarray}
since $b_{q}|ph,0\left.\right\rangle=0$. As the band is half-filled the Fermi energy 
$\epsilon_{F}=0$. Thus $c^{\dag}_{k+q}c_{k}|FS\left.\right\rangle\not=0$ only if 
$\epsilon_{k+q}\ge 0$ and $\epsilon_{k}\le 0$. So, we have  
\begin{eqnarray}
(H^{1}+[S,H^{0}])|g_{0}\left.\right\rangle & = & 0,
\end{eqnarray}
if we choose
\begin{equation}
\delta(k+q,k)=1/(1+|\epsilon_{k+q}-\epsilon_{k}|/\omega_{\pi}).
\end{equation}
This is nothing but making the matrix element of $H^{1}+[S,H^{0}]$between
$|g_{0}\left.\right\rangle$ and the lowest-lying excited states vanishing.
Thus the first-order terms, which are not exactly canceled after the
transformation, are related to the higher-lying excited states and should be
irrelevant under renormalization\cite{refa14}.The second-order terms in $H'$
can be collected as follows:
\begin{eqnarray}
  &   & [S,H^{1}]+\frac{1}{2}[S,[S,H^{0}]]\nonumber\\ 
  & = & \frac{1}{2N}\sum_{q,k,s}\sum_{q',k'}\frac{g(q)g(q')}{\omega_{\pi}}
        \delta(k+q,k)[2-\delta(k'+q',k')](b^{\dag}_{-q'}+b_{q'})
        (b^{\dag}_{-q}-b_{q})\nonumber\\                
  & \times & (c^{\dag}_{k+q,s}c_{k',s}\delta_{k,k'+q'}
        -c^{\dag}_{k'+q',s}c_{k,s}\delta_{k',k+q})\nonumber\\
  & + & \frac{1}{2N}\sum_{q,k,s}\sum_{q',k'}\frac{g(q)g(q')}
        {\omega_{\pi}^{2}}\delta(k+q,k)\delta(k'+q',k')(\epsilon_{k'}-
        \epsilon_{k'+q'})(b^{\dag}_{-q'}-b_{q'})
        (b^{\dag}_{-q}-b_{q})\nonumber\\                
  & \times & (c^{\dag}_{k+q,s}c_{k',s}\delta_{k,k'+q'}
        -c^{\dag}_{k'+q',s}c_{k,s}\delta_{k',k+q})\nonumber\\
  & - & \frac{1}{2N}\sum_{q,k,s}\sum_{k',s'}
        \frac{g(q)g(-q)}{\omega_{\pi}}
        \delta(k+q,k)[2-\delta(k'-q,k')]c^{\dag}_{k+q,s}c_{k,s}
        c^{\dag}_{k'-q,s'}c_{k',s'}.
\end{eqnarray}
$\delta_{k',k+q}$ is the Kronecker $\delta$ symbol. All terms of higher 
order than $\alpha^{2}$ will be omitted in the following treatment.
  
  For the dimerized state, the neighboring atoms move in opposite directions.
 To take into account the static phonon-staggered ordering\cite{refa6}, we
 make a displacement transformation to $H'$  
\begin{equation}
\tilde{H}=\exp(R)H'\exp(-R).
\end{equation}
Here
\begin{equation}
 R=-\sum_{l}(-1)^{l}u_{0}\sqrt{\frac{M\omega_{\pi}}{2}}(b^{\dag}_{l}-b_{l}),
\end{equation}
and $\exp(R)$ is a displacement operator:
\begin{equation}
 \exp(R)u_{l}\exp(-R)=(-1)^{l}u_{0}+\sum_{q}\sqrt{\frac{1}{2MN\omega_{q}}}
                      (b^{\dag}_{-q}+b_{q})\exp(iql).
\end{equation}
Applying the transformation on phonon annihilation operator, we get  
\begin{equation}
 \exp(R)b_{q}\exp(-R)=b_{q}+u_{0}\sqrt{\frac{NM\omega_{\pi}}{2}}\delta_{\pi,q}.
\end{equation}
If the ground state of $H$ is $|g\left.\right\rangle$, then the ground state
of $\tilde{H}$ is $|g'\left.\right\rangle$:
$|g\left.\right\rangle=\exp(-S)\exp(-R)|g'\left.\right\rangle$. We assume
that for $|g'\left.\right\rangle$ the fermions and phonons can be decoupled:
$|g'\left.\right\rangle\approx |fe\left.\right\rangle|ph,0\left.\right\rangle
$, where $|fe\left.\right\rangle$ is the ground state for fermions. After
averaging $\tilde H$ over the phonon vacuum state we get an effective
Hamiltonian for the fermions,
\begin{eqnarray}
 H_{eff} & = & \left\langle ph,0|\tilde{H}|ph,0\right\rangle\nonumber\\
         & = & \frac{1}{2}Ku_{0}^{2}N+\sum_{k,s}E_{0}(k)c^{\dag}_{k,s}c_{k,s}
               +\sum_{k>0,s}\Delta_{0}(k)(c^{\dag}_{k-\pi,s}c_{k,s}
               +c^{\dag}_{k,s}c_{k-\pi,s})\nonumber\\
         & - & \frac{1}{N}\sum_{q,k,s}\sum_{k',s'}
               \frac{g(q)g(-q)}{\omega_{\pi}}
               \delta(k+q,k)[2-\delta(k'-q,k')]c^{\dag}_{k+q,s}c_{k,s}
               c^{\dag}_{k'-q,s'}c_{k',s'},
\end{eqnarray}
where
\begin{eqnarray}
E_{0}(k) & = & \epsilon_{k}-\frac{1}{N}\sum_{k'}
               \frac{g(k'-k)g(k-k')}{\omega^{2}_{\pi}}\delta(k',k)\delta(k,k')
                 (\epsilon_{k}-\epsilon_{k'}),\\
\Delta_{0}(k) & = & 2\alpha u_{0}[1-\delta(k-\pi,k)].
\end{eqnarray}
We find by means of the variational principle that the dimerized lattice 
displacement ordering parameter is\\
\begin{equation}
 u_{0}=-\frac{1}{KN}\sum_{k>0,s}2\alpha [1-\delta(k-\pi,k)]
       \left\langle fe\left|(c^{\dag}_{k-\pi,s}c_{k,s}
       +c^{\dag}_{k,s}c_{k-\pi,s})\right|fe\right\rangle.
\end{equation}
  
  Note that in the adiabatic limit where $\omega_{\pi}=0$ one has 
$\delta(k',k)=0$ and $H_{eff}$ goes back to the adiabatic mean-field 
Hamiltonian, 
\begin{equation}
 H_{eff}(\omega_{\pi}=0)=\frac{1}{2}KNu_{0}^{2}+\sum_{k,s}\epsilon_{k}
                        c^{\dag}_{k,s}c_{k,s}+\sum_{k>0,s}2\alpha u_{0}
                        (c^{\dag}_{k-\pi,s}c_{k,s}+c^{\dag}_{k,s}c_{k-\pi,s}).
\end{equation}
This is the same as the adiabatic mean-field Hamiltonian for the dimerized
Holstein model\cite{refa14}.
On the other hand, in the antiadiabatic limit where $\omega_{\pi}\rightarrow
\infty$, we have $u_{0}=0$, $\delta(k',k)=1$ , and $H_{eff}$ becomes  
\begin{equation}
 H_{eff}(\omega_{\pi}\rightarrow\infty)=\sum_{k,s}\epsilon_{k}
               c^{\dag}_{k,s}c_{k,s}-\frac{1}{N} 
               \sum_{k,k'}\sum_{q,s,s'}\frac{g(q)g(-q)}{\omega_{\pi}}
               c^{\dag}_{k+q,s}c_{k,s}c^{\dag}_{k'-q,s'}c_{k',s'}.
\end{equation}
Returning to the real space, this Hamiltonian is
\begin{equation}
 H_{eff}(\omega_{\pi}\rightarrow\infty)=-t_{0}\sum_{l,s}
               (c^{\dag}_{l+1,s}c_{l,s}+c^{\dag}_{l,s}c_{l+1,s})
               -\frac{\alpha^{2}}{K}\sum_{l,s,s'}
               c^{\dag}_{l,s}c_{l,s}c^{\dag}_{l,s'}c_{l,s'}
               +\frac{\alpha^{2}}{K}\sum_{l,s,s'}
               c^{\dag}_{l,s}c_{l,s}c^{\dag}_{l+1,s'}c_{l+1,s'}.
\label{ome}
\end{equation}
For the spinless case, the on-site interaction disappears because of the Pauli
principle and this is the antiferromagnetic XXZ model (through
Jordan-Wigner trans\-for\-mation)\cite{refa5}. The situation is very similar
to the spinless SSH mode in the large $\omega_{\pi}$ limit \cite{refa19}.
XXZ model can be solved exactly and
there exists a transition point at $\alpha^{2}/2K=t_{0}$\cite{refa15}.
For the spin-{\footnotesize $\frac{1}{2}$} case, (\ref{ome}) is the
negative-$U$ extended Hubbard model with $V=-U$.

Our effective Hamiltonian works well in the $\omega_{\pi}=0$ and
$\omega_{\pi}\rightarrow\infty$ limits and, furthermore, for the spinless
case $H_{eff}$ can be solved exactly in both limits. In this work we
concentrate on the nonadiabatic effect due to finite phonon frequency
$\omega_{\pi}$ in the spinless case because we have the exact solutions in
the two limits. The last term in $H_{eff}$ is a four-fermion interaction.
As we are dealing with a one-dimensional system, how to treat
the four-fermion interaction is
a difficult problem. Since the case for the small $\omega_{\pi}$, 
$\omega_{\pi}\le 2t_{0}$, is very different from that for the large  
$\omega_{\pi}$, $\omega_{\pi}>2t_{0}$, we treat $H_{eff}$ in these two 
cases with different methods. 
\section{$ \omega_{\pi}<2t_{0} $}
  In this case the four-fermion term goes to zero as $\omega_{\pi}
\rightarrow0$, so it can be treated as a perturbation and the
unperturbed Hamiltonian is
\begin{eqnarray}
 H_{eff}^{0} & = & \frac{1}{2}Ku_{0}^{2}N+\sum_{k}E_{0}(k)c^{\dag}_{k}c_{k}
                   +\sum_{k>0}\Delta_{0}(k)
                   (c^{\dag}_{k-\pi}c_{k}+c^{\dag}_{k}c_{k-\pi}).
\end{eqnarray}
The four-fermion term can be re-written as
\begin{eqnarray}
 H'_{eff} & = & \frac{1}{N}\sum_{q>0,}\sum_{k>0,k'>0}
                \frac{g(q)g(-q)}{\omega_{\pi}}\delta(k+q,k)
                [2-\delta(k'-q,k')]\nonumber\\
          &  & \times (c^{\dag}_{k+q}c_{k}+c^{\dag}_{k+q-\pi}c_{k-\pi})
                      (c^{\dag}_{k'-q}c_{k'}+c^{\dag}_{k'-q-\pi}
                      c_{k'-\pi})\nonumber\\
          &  & -\frac{1}{N}\sum_{q>0,}\sum_{k>0,k'>0}
                \frac{g(q-\pi)g(\pi-q)}{\omega_{\pi}}\delta(k+q,k-\pi)
                [2-\delta(k'-q,k'-\pi)]\nonumber\\
          &  & \times(c^{\dag}_{k+q}c_{k-\pi}c^{\dag}_{k'-q-\pi}c_{k'}
                      +c^{\dag}_{k+q-\pi}c_{k}c^{\dag}_{k'-q}
                      c_{k'-\pi})\nonumber\\
          &  & -\frac{1}{N}\sum_{q>0,}\sum_{k>0,k'>0}
                \frac{g(q-\pi)g(\pi-q)}{\omega_{\pi}}\delta(k+q,k-\pi)
                [2-\delta(k'-q,k'-\pi)]\nonumber\\
          &  & \times(c^{\dag}_{k+q-\pi}c_{k}c^{\dag}_{k'-q-\pi}c_{k'}
                      +c^{\dag}_{k+q}c_{k-\pi}c^{\dag}_{k'-q}
                      c_{k'-\pi}).
\label{eq:Hamil}
\end{eqnarray}
We can distinguish between different physical processes. The first term in 
Eq.(\ref{eq:Hamil}) is the forward scattering one, the second is the backward  
scattering one, and the last is the Umklapp scattering one. We use the Green's
function method to implement the perturbation treatment and it is more
convenient to work within a two-component representation,
\begin{eqnarray}
 \Psi_{k} & = & \left({\begin{array}{l}
                    c_{k}\\
                    c_{k-\pi}
             \end{array}}
      \right)
             ,\ \ \ \ \ \ \ \ \ k>0.
\end{eqnarray}
Thus we have
\begin{eqnarray}
  \left\{{\begin{array}{l}
             \Psi^{\dag}_{k}\sigma_{z}\Psi_{k}
                      =c_{k}^{\dag}c_{k}-c_{k-\pi}^{\dag}c_{k-\pi}\\
             \Psi^{\dag}_{k}\sigma_{x}\Psi_{k}
                      =c_{k}^{\dag}c_{k-\pi}+c_{k-\pi}^{\dag}c_{k}\\       
             \Psi^{\dag}_{k}i\sigma_{y}\Psi_{k}
                      =c_{k}^{\dag}c_{k-\pi}-c_{k-\pi}^{\dag}c_{k}\\
             \end{array}}
      \right.,
\end{eqnarray}
and the Hamiltonian becomes
\begin{eqnarray}
 H_{eff}^{0} & = & \frac{1}{2}Ku_{0}N+\sum_{k>0}E_{0}(k)
                    \Psi^{\dag}_{k}\sigma_{z}\Psi_{k}
                   +\sum_{k>0}\Delta_{0}(k)
                   \Psi^{\dag}_{k}\sigma_{x}\Psi_{k},
\end{eqnarray}
\begin{eqnarray}
 H'_{eff} & = & -\frac{1}{N}\sum_{q>0,}\sum_{k>0,k'>0}
                \frac{g(q)g(-q)}{\omega_{\pi}}\delta(k+q,k)
                [2-\delta(k'-q,k')]
                \Psi^{\dag}_{k+q}\Psi_{k}\Psi^{\dag}_{k'-q}\Psi_{k'}
                \nonumber\\
          &  & +\frac{1}{2N}\sum_{q>0,}\sum_{k>0,k'>0}
                \frac{g(q-\pi)g(\pi-q)}{\omega_{\pi}}
                \delta(k+q,k-\pi)[2-\delta(k'-q,k'-\pi)]\nonumber\\
          &  & \times(\Psi^{\dag}_{k+q}i\sigma_{y}\Psi_{k}
                       \Psi^{\dag}_{k'-q}i\sigma_{y}\Psi_{k'}
                       -\Psi^{\dag}_{k+q}\sigma_{x}\Psi_{k}
                       \Psi^{\dag}_{k'-q}\sigma_{x}\Psi_{k'})\nonumber\\
          &  & -\frac{1}{2N}\sum_{q>0,}\sum_{k>0,k'>0}
                      \frac{g(q-\pi)g(\pi-q)}{\omega_{\pi}}
                      \delta(k+q,k-\pi)[2-\delta(k'-q,k'-\pi)]\nonumber\\
          &  & \times(\Psi^{\dag}_{k+q}i\sigma_{y}\Psi_{k}
                      \Psi^{\dag}_{k'-q}i\sigma_{y}\Psi_{k'}+
                     \Psi^{\dag}_{k+q}\sigma_{x}\Psi_{k}
                     \Psi^{\dag}_{k'-q}\sigma_{x}\Psi_{k'}).
\end{eqnarray}
$\sigma_{\beta}(\beta=x,y,z)$ is the Pauli matrix. The matrix Green's function 
is defined as (the temperature Green's function is used and at the end let 
$T\rightarrow0$)
\begin{eqnarray}
 G(k,\tau) & = & -<T_{\tau}\Psi_{k}(\tau)\Psi^{\dag}_{k}(0)>\nonumber\\
           & = & \frac{1}{\beta}\sum_{n}\exp(-i\omega_{n}\tau)G(k,\omega_{n}).
\end{eqnarray}
The Dyson equation is
\begin{eqnarray}
 G(k,\omega_{n}) & = & G_{0}(k,\omega_{n})+G_{0}(k,\omega_{n})
                       \Sigma ^{*}(k,\omega_{n})G(k,\omega_{n}),
\end{eqnarray}
where
\begin{eqnarray}
 G_{0}(k,\omega_{n}) & = & \left\{ i\omega_{n}-E_{0}(k)\sigma_{z}
                         -\Delta_{0}(k)\sigma_{x} \right\} ^{-1},
\end{eqnarray}
is the unperturbed Green's function. The self-energy 
$\Sigma ^{*}(k,\omega_{n})$ can be calculated by the perturbation theory,
\begin{eqnarray}
 \Sigma ^{*}(k,\omega_{n})\nonumber& = & \frac{T}{N}\sum_{k'}\sum_{m}
                \frac{g(k'-k)g(k-k')}{\omega_{\pi}}\delta(k',k)
                [2-\delta(k,k')]\nonumber\\
  & \times & \left\{G_{0}(k',\omega_{m})+T_{r}[\sigma_{z}
                      G_{0}(k',\omega_{m})]\sigma_{z}\right\}\nonumber\\
  & - & \frac{T}{N}\sum_{k'}\sum_{m}
            \frac{g(k'-k-\pi)g(\pi-k'+k)}{\omega_{\pi}}
            \delta(k',k-\pi)[2-\delta(k,k'-\pi)]\nonumber\\
  & \times & \left[i\sigma_{y}G_{0}(k',\omega_{m})i\sigma_{y}-
            \sigma_{x}G_{0}(k',\omega_{m})\sigma_{x}\right]\nonumber\\
  & + & \frac{T}{N}\sum_{k'}\sum_{m}
           \frac{g(-\pi)g(\pi)}{\omega_{\pi}}
           [\delta(k,k-\pi)+\delta(k',k'-\pi)-
           \delta(k,k-\pi)\delta(k',k'-\pi)]\nonumber\\
  & \times & \left\{T_{r}[i\sigma_{y}G_{0}(k',\omega_{m})]i\sigma_{y}+
              T_{r}[\sigma_{x}G_{0}(k',\omega_{m})]\sigma_{x}\right\}.
\label{eq:sener}
\end{eqnarray}
In the perturbation calculation we have taken into account the fact that the 
forward and backward scattering terms contribute nothing to the "charge" gap
\cite{refa17,refa18}. From Eq.(\ref{eq:sener}) one can get that
$\Sigma ^{*}(k,\omega_{n})$ is
irrelative to $\omega_{n}$, therefore the spectrum structure of 
$G(k,\omega_{n})$ should be
\begin{eqnarray}
G(k,\omega_{n}) & = & \left\{ i\omega_{n}-E(k)\sigma_{z}
                        -\Delta(k)\sigma_{x} \right\} ^{-1}.
\end{eqnarray} 
From $G(k,\omega_{n})$ the fermionic spectrum in the gapped state can be 
derived 
\begin{eqnarray}
W(k) & = & \sqrt{E^{2}(k)+\Delta^{2}(k)}.
\end{eqnarray} 
The renormalized band function and the gap function are
\begin{eqnarray}
 E(k) & = & E_{0}(k)-\frac{2\alpha^{2}}{KN}\sum_{k'>0}
            \left\{\sin^{2}\left(\frac{k'-k}{2}\right)\delta(k',k)
            [2-\delta(k',k)]\right. \nonumber\\
      &  & -\left. \cos^{2}\left(\frac{k'-k}{2}\right)\delta(k'-\pi,k)
            [2-\delta(k'-\pi,k)] \right\}
            \frac{E_{0}(k')}{\sqrt{E_{0}^{2}(k')+\Delta^{2}_{0}(k')}},\\
\Delta(k) & = & 2\alpha u_{0}[c-d\delta(k-\pi,k)].
\label{delta}
\end{eqnarray}
Where
\begin{eqnarray}
 c & = & 1+\frac{2\alpha^{2}}{KN}\sum_{k>0}\delta(k-\pi,k)
         \frac{\Delta_{0}(k)}{2\alpha u_{0}\sqrt{E_{0}^{2}(k)
         +\Delta^{2}_{0}(k)}},\\
 d & = & 1-\frac{2\alpha^{2}}{KN}\sum_{k>0}[1-\delta(k-\pi,k)]
         \frac{\Delta_{0}(k)}
         {2\alpha u_{0}\sqrt{E_{0}^{2}(k)+\Delta^{2}_{0}(k)}},
\end{eqnarray}
The equation to determine $u_{0}$ is
\begin{equation}
 1=\frac{4\alpha^{2}}{KN}\sum_{k>0}[1-\delta(k-\pi,k)]
         \frac{\Delta(k)}{2\alpha u_{0}W(k)}.
\label{uofun}
\end{equation}
In nonadiabatic case $u_{0}$ is a variational parameter and cannot be 
measured by experiment or Monte Carlo simulation. The quantity which can be 
measured is the phonon-staggered ordering parameter $m_{p}$,
\begin{eqnarray}
 m_{p} & = & \frac{1}{N}\sum_{l}(-1)^{l}<u_{l}>\nonumber\\
       & = & \frac{1}{N}\sum_{l}\sum_{q}(-1)^{l}
             \sqrt{\frac{1}{2MN\omega_{\pi}}}\exp(iql)
             <b^{\dag}_{-q}+b_{q}>\nonumber\\
       & = & u_{0}-\frac{2\alpha}{KN}\sum_{k}\delta(k-\pi,k)
             <c^{\dag}_{k-\pi}c_{k}>\nonumber\\  
       & = & \frac{2\alpha}{KN}\sum_{k>0}<fe|\Psi^{\dag}_{k}\sigma_{x}
             \Psi_{k}|fe>\nonumber\\
       & = & \frac{2\alpha}{KN}\sum_{k>0}\frac{\Delta(k)}{W(k)}.
\end{eqnarray}
These are basic equations for the $\omega_{\pi}<2t_{0}$ case. If  
$\omega_{\pi}=0$ we have $\delta(k',k)=0$ and $c=1$, Eq.(\ref{uofun}) becomes 
the same as that in the adiabatic theory.

Fig.1 shows the phonon-staggered ordering parameter
$\alpha m_{p}/t_{0}$ as function of the electron-phonon coupling constant
$\alpha^{2}/Kt_{0}$ in the cases of $\omega_{\pi}/t_{0}=0.3$ and 0.5.
As shown in the figure, the dimerization parameter $\alpha m_{p}/t_{0}$
increases as the electron-phonon coupling constant $\alpha^{2}/Kt_{0}$
increases but decreases as the phonon frequency $\omega_{\pi}/t_{0}$
increases. Fig.2 shows the normalized phonon-staggered ordering
parameter $m_{p}/m_{p0}$ ($m_{p0}$ is the adiabatic value when
$\omega_{\pi}=0$) as functions of the normalized phonon frequency
$\omega_{\pi}/\omega_{\pi c}$ ($\omega_{\pi c}$ is the value when
$m_{p}=0$) in the cases of $\alpha^{2}/Kt_{0}=0.5$ and 1.0. As shown
in the figure, the dimerization
parameter $m_{p}$ decreases as the phonon frequency $\omega_{\pi}$ increases
but increases as the electron-phonon coupling constant $\alpha^{2}/Kt_{0}$
increases, which indicates that the nonadiabatic effect is to reduce the
dimerization ordering parameters.

  Comparing Eq.(\ref{delta}) with that in the adiabatic case, $\Delta(k)=
2\alpha u_{0}$, we have the gap in the nonadiabatic case,
\begin{equation}
 \Delta=\Delta(\pi/2)=2\alpha u_{0}[c-d].
\end{equation}
This is the true gap in the fermionic spectrum.

  From Eq.(\ref{uofun}), let $u_{0}=0$, we get the self-consistent equation 
of phase transition points in the $\alpha^{2}/K\sim\omega_{\pi}$ plane,
\begin{equation}
 1=\frac{4\alpha^{2}}{KN}\sum_{k>0}[1-\delta(k-\pi,k)]
         \frac{c-d\delta(k-\pi,k)}{|E(k)|}.
\label{figu41}
\end{equation}

  The density of states(DOS) of fermions is
\begin{eqnarray}
 N(\omega) & = & \frac{1}{N}\sum_{k}\delta\left(\omega-
                 \sqrt{E^{2}(k)+\Delta^{2}(k)}\right)\nonumber\\
         & = & \frac{1}{2\pi}\left(\frac{d}{dk}
             \sqrt{E^{2}(k)+\Delta^{2}(k)}
             \left|_{k=f(\omega)}\right.\right)^{-1},
\end{eqnarray}
where, $k=f(\omega)$ is the inverse function of 
$\omega=\sqrt{E^{2}(k)+\Delta^{2}(k)}$. Fig.3 shows the density of states(DOS)
of fermions for some values of phonon frequency $\omega_{\pi}$. One can see
that a nonzero
DOS starts from the gap edge and, for small values of $\omega_{\pi}$, there
is a peak with a significant tail below it. The inverse-square-root
singularity at the gap edge in the adiabatic case\cite{refa19} disappears.
\section{$ \omega_{\pi}>2t_{0} $}
  In this case $H_{eff}$ can be re-written as
\begin{eqnarray}
 H_{eff} & = & \frac{1}{2}Ku_{0}^{2}N+\sum_{k}E_{0}(k)c^{\dag}_{k}c_{k}
               +\sum_{k>0}\Delta_{0}(k)
               (c^{\dag}_{k-\pi}c_{k}+c^{\dag}_{k}c_{k-\pi})\nonumber\\
         & - & \frac{1}{N}\sum_{q,k,k'}\frac{g(q)g(-q)}{\omega_{\pi}}
               \delta(k+q,k)[2-\delta(k'-q,k')]
               c^{\dag}_{k+q}c_{k}c^{\dag}_{k'-q}c_{k'}.
\label{hhh}
\end{eqnarray}

  According to Eq.(\ref{ome}), when $\omega_{\pi}\rightarrow\infty$, the 
effective Hamiltonian includes only electron-electron interactions of on-site        
and nearest-neighbor. Therefore, we treat the terms of on-site and
nearest-neighbor 
interactions as the unperturbed Hamiltonian and the others as perturbation 
because they go to zero when $\omega_{\pi}\rightarrow\infty$. In real space,
the unperturbed Hamiltonian is
\begin{eqnarray}
 H^{0}_{eff} & = & \frac{1}{2}Ku_{0}^{2}N-t_{0}[1-F(\omega_{\pi})]
                   \sum_{l}(c^{\dag}_{l}c_{l+1}+c^{\dag}_{l+1}c_{l})
                   +V_{1}\sum_{l}c^{\dag}_{l}c_{l}c^{\dag}_{l+1}c_{l+1},
\label{llll}
\end{eqnarray}
where
\begin{eqnarray}
 V_{1} & = & -\frac{2}{N^{3}}\sum_{q,k,k'}\frac{|g(q)|^{2}\cos q}{\omega_{\pi}}
              \delta(k+q,k)[2-\delta(k'-q,k')],
\end{eqnarray}
\begin{eqnarray}
 F(\omega_{\pi})&= & \frac{\omega_{\pi}}{(2\pi)^{2}}\frac{2\alpha^{2}}{K}
                       \int_{-\pi}^{\pi}dk\int_{-\pi}^{\pi}dk'
                       \frac{(1-\cos k \cos k')(\cos k-\cos k')\cos k}
                       {(\omega_{\pi}+2t_{0}|\cos k-\cos k'|)^{2}}.
\end{eqnarray}
(\ref{llll}) is the antiferromagnetic XXZ model and can be solved exactly
\cite{refa5}. The result of Yang-Yang shows that there exists a transition
point at\cite{refa15}
\begin{eqnarray}
 V_{1} &= & 2t_{0}[1-F(\omega_{\pi})].
\label{figu42}
\end{eqnarray}
This equation determines the diagram of phase transition in the case of 
$\omega_{\pi}>2t_{0}$. Fig.4 shows the phase diagram. We use $\omega_{\pi}
/(\omega_{\pi}+2t_{0})$, instead of $\omega_{\pi}$, as the variable because it
goes to 1 when $\omega_{\pi}\rightarrow\infty$. The solid line is the result
of Eq.(\ref{figu41}) and the dashed line is the result obtained from
Eq.(\ref{figu42}). The dashed-dotted line is the result of equation
\begin{eqnarray}
 \frac{\alpha^{2}}{K} & \sim & \left(\frac{\omega_{\pi}}{\omega_{\pi}+2t_{0}}
                               \right)^{0.4}.
\label{figu43}
\end{eqnarray}
One can see that, although the formula is
very simple, the interpolated result is, at least, qualitatively correct.  
\section{Conclusions}
  The effects of quantum lattice fluctuations on the ground state of 
a model electron-phonon system are studied through an analytical approach.
Our results show that when the
electron-phonon coupling constant $\alpha^{2}/K$ decreases or the phonon 
frequency $\omega_{\pi}$ increases the lattice dimerization and the gap in 
the fermion spectrum decrease gradually. At some critical value, the system 
becomes gapless and the lattice dimerization disappears. A phase diagram in
the $\alpha^{2}/K\sim\omega_{\pi}$ plane and the density
of states of fermions are derived. The inverse-square-root singularity of DOS 
at the gap edge in the adiabatic case disappears because of the nonadiabatic 
effect, which is consistent with the measurement of optical conductivity in
quasi-one-dimensional systems.

\newpage
{\bf\Large Figure Caption}\\
\newline
{\bf Fig.1} The dimerization parameter $\alpha m_{p}/t_{0}$ as function of the
electron-phonon coupling constant $f=\alpha^{2}/Kt_{0}$ for
$\omega_{\pi}/t_{0}=0.3$ and 0.5.\\
\newline
{\bf Fig.2} The normalized dimerization parameter $m_{p}/m_{p0}$ as functions
of the normalized phonon frequency $\omega_{\pi}/\omega_{\pi c}$ in the
cases of $f=0.5$ and 1.0.\\
\newline
{\bf Fig.3} The density of states(DOS) of fermions for
$\alpha^{2}/Kt_{0}=0.4$ with $\omega_{\pi}/t_{0}=0.001$ and 0.01.\\
\newline
{\bf Fig.4} The phase diagram from Eq.(\ref{figu41}), Eq.(\ref{figu42}) and
Eq.(\ref{figu43}).

\end{document}